\input{aipcheck}
\documentclass[
    ,final            
  ]
  {aipproc}

\layoutstyle{6x9}

\usepackage{amsmath}

\newcommand{\yr}{{\,\rm yr}}

\newcommand{\Myr}{{\,\rm Myr}}
\newcommand{\perGyr}{{\,\rm Gyr^{-1}}}
\newcommand{\pc}{\,\mathrm{pc}}
\newcommand{\Gpc}{\,\mathrm{Gpc}}

\newcommand{\Mbh}{M_{\bullet}}

\newcommand{\Mo}{M_{\odot}}

\newcommand{\Ms}{M_{\star}}
\newcommand{\ns}{n_{\star}}

\newcommand{\LISA}{{\it LISA\,}}

\newcommand{\apj}{ApJ}

\newcommand{\apjl}{ApJL}

\newcommand{\mnras}{MNRAS}
\newcommand{\aap}{A.A.P.}
\newcommand{\physrep}{Phys. Rep.}
\newcommand{\nat}{Nature}

\begin{document}
\bibliographystyle{apj.bst} 
%

%
%

\title{Astrophysics of extreme mass ratio inspiral sources}



\author{Clovis Hopman} {address={Faculty of Physics, Weizmann
Institute of Science, POB 26, Rehovot 76100, Israel; Leiden
Observatory, P.O. Box 9513, NL-2300 RA Leiden.
{\tt E-mail:}clovis@strw.leidenuniv.nl}}


\begin{abstract}
Compact remnants on orbits with peri-apses close to the Schwarzschild
radius of a massive black hole (MBH) lose orbital energy by emitting
gravitational waves (GWs) and spiral in. Scattering with other stars
allows successful inspiral of such extreme mass ratio inspiral sources
(EMRIs) only within small distances, {{{}}}$a<\,\,{\rm
few}\times0.01\pc$ from the MBH. The event rate of EMRIs is therefore
dominated by the stellar dynamics and content in the inner ${\rm
few}\times0.01\pc$. I discuss the relevant dynamical aspects and
resulting estimated event rates of EMRIs. Subjects considered include
the loss-cone treatment of inspiral sources; mass segregation;
resonant relaxation; and alternative routes to EMRI formation such as
tidal binary disruptions, stellar formation in disks and tidal capture
of massive main sequence stars. The EMRI event rate is estimated to be
of order ${\rm few}\times10^2\perGyr$ per MBH, giving excellent
prospects for observation by \LISA.
\end{abstract}

\maketitle

\section{Introduction}

One of the most exciting and plausible targets of the {\it Laser
Interferometer Space Antenna} (\LISA) is the inspiral of a stellar
mass compact remnant (CR), i.e., a white dwarf (WD), neutron star (NS)
or stellar black hole (BH), into a massive black hole (MBH) of mass
$\Mbh\sim10^6\Mo$, or an intermediate mass black hole (IMBH) of
$\Mbh\sim(10^3-10^4)\Mo$. Because of the large difference between the
mass of the MBH and the CR, such sources are known as ``extreme mass
ratio inspirals'' (EMRIs). \LISA can detect stars with orbital periods
$1\,{\rm s}<P<10^4\,{\rm s}$ at distances of $\sim1\Gpc$.

MBHs are embedded in dense stellar cusps, as was predicted
theoretically \citep{Bah76, Bah77}, and confirmed observationally in
our own Galactic center
\citep{Ale99a, Gen03a,Ale05}. Due to the high stellar density, the
relaxation time, i.e., the time it takes a star to change its energy
by order unity, is typically shorter than the age of the
Universe. This implies that the orbits of stars can change due to
dynamical encounters in the course of time: a star that is initially
not on an orbit at which it loses much energy to gravitational waves
(GWs), can be scattered to an orbit where energy losses to GWs are
large, so that the star spirals in.

The short relaxation time of galactic nuclei complicates the analysis
of EMRI rates, as is reflected by wildly differing estimates in the
literature \citep{Hil95, Sig97b, Mir00, Fre01, Iva02, Fre03, Ale03b,
Gai04, Hop05, Hop06, Hop06b}. Here I discuss a framework in which EMRI
dynamics can be analyzed. The most important result is that the EMRI
rate is mostly determined by the dynamics and stellar content of the
inner few 0.01 pc. Within this distance from the MBH, dynamical
effects such as mass segregation and resonant relaxation are
discussed. I also review some alternative EMRI formation scenarios,
which differ from the direct capture model.

\section{Loss-cone dynamics of EMRIs}\label{s:losscone}

\subsection{Stars near a MBH: energy space}

The distribution function (DF) of Newtonian point masses near a MBH was
first found by Bahcall \& Wolf \citep{Bah76} for a single mass
population, and later generalized to a multi-mass population
\citep{Bah77}. The Bahcall \& Wolf solution is characterized by a
density profile of stars that diverges towards the center. For a
single mass population, the density is given by

\begin{equation}\label{e:cusp}
\ns\propto r^{-\alpha},
\end{equation}
with $\alpha=7/4$. For a multi-mass population, the slope $\alpha$
depends on mass. First, I focus on the single mass case, which
captures much of the relevant physics. I later return to the important
(but not much studied) case of a multi-mass population.

An interesting feature of the Bahcall \& Wolf solution is that it is
not only a steady state solution of the Fokker-Planck equation, but
that in addition the flow of stars in energy space is very strongly
suppressed compared to what one would expect from dimensional
arguments. As a result of this, stars typically come close to the MBH
due to scattering in their angular momenta rather than in their
energies.

\subsection{Stars near a MBH: angular momentum space}

Stars typically reach the MBH on wide orbits of low energy that are
very eccentric (have very small angular momentum) \citep{Lig77,
Bah77}. If there is some periapse $r_q$ such that stars with periapse
$r_p<r_q$ are swallowed by the MBH, then this defines for loosely
bound orbits a region in angular momentum space called the
``loss-cone'', given by $J_{lc}=\sqrt{2G\Mbh r_{q}}$. For CRs, the
loss-cone is determined by the angular momentum of the last stable
orbit in general relativity, which for an eccentric orbit is given by
$J_{lc}=J_{\rm LSO}\equiv 4G\Mbh/c$.

Far away from the MBH, the changes $\Delta J_p$ in angular momentum
per orbit are much larger than the size of the loss-cone. This implies
that a star for which $J<J_{lc}$ at apo-apse may have $J>J_{lc}$ at
peri-apse and survive, and vise-versa: a star with $J>J_{lc}$ at
apo-apse may have $J<J_{lc}$ at peri-apse and be swallowed. As a
result, far away from the MBH, the DF can be isotropic, and for given
semi-major axis the fraction of stars with angular momentum $<J$ is
given by $(J/J_c)^2$, where $J_c$ is the angular momentum of a
circular orbit. The rate at which stars are swallowed by the MBH in
this ``full loss-cone'' regime is estimated by $\Gamma(a)da\sim
N(a)da(J_{lc}/J_c)^2/P(a)$, where $N(a)da$ is the number of stars in
the interval $(a, a+da)$.

Close to the MBH, $\Delta J_p\ll J_{lc}$, so that any star in the
loss-cone is immediately swallowed by the MBH. Since the time-scale
for a change in angular momentum of order $J_c$ is the relaxation time
$t_r$, the rate at which stars are captured by the MBH in this ``empty
loss-cone'' regime is
\begin{equation}
\Gamma(a)da=\frac{N(a)}{\ln(J_{c}/J_{lc})t_{r}(a)}da\,.\label{e:Flow}
\end{equation}
The logarithmic factor in the denominator stems from a modification of
the DF due to the presence of the loss-cone \citep{Lig77}. Since
inspiral occurs over many orbits, it is the empty loss-cone regime
which is relevant for the GW event rate.

\subsection{Inspiral sources}\label{ss:insp}

Equation (\ref{e:Flow}) assumes that a star that is captured at
peri-apse is taken out of the system immediately. Such events indeed
occur (for example when a CR has a wide orbit with peri-apse smaller
than the Schwarzschild radius), but such stars do not contribute to
the LISA detection rate\footnote{Our own Galactic center may be close
enough to observe bursts of GWs when stars come close to the MBH at
one single passage \citep{Rub06}. In the following I do not consider
such events, which are not observable in extra-galactic nuclei.}. In
order to be detected by \LISA, a star may start at a wide orbit, but
must then spiral in until its period becomes comparable to
{{{}}}$P<10^4$ s.

A star that starts spiraling in when its semi-major axis is very
large, is unlikely to reach the MBH without plunging in, because it
has to make a very large number of orbits. Since during these orbits
its angular momentum changes randomly due to weak encounters with
other stars, the probability is large that it will at some point be
scattered into the MBH. On the other hand, a star that has initially a
small semi-major axis, is likely to spiral in without being scattered
into the loss-cone, since it loses energy quickly and has little
chance to scatter significantly. Thus, there exists some critical
semi-major axis $a_{\rm GW}$ that separates these two regimes. This
critical value can be estimated as follows \citep{Hop05}:

The change in $J$ of a star per orbital period $P$ is $\Delta J_p =
(P/t_r)^{1/2}J_c$. The time-scale for a change of order $J$ is
$t_J=(J/J_c)^2t_r$. In particular, the time-scale for a change in $J$
by the order of the loss-cone, is $t_{lc}=(J_{\rm
LSO}/J_c)^2t_r$. Inspiral due to dissipation by GW emission happens on
a time-scale $t_0(a,J)$, which for highly eccentric orbits has a very
strong angular momentum dependence, $t_0(J)\propto J^{7}$. If
$t_{lc}\ll t_0(a, J\!\to\!  J_{\rm LSO})$, the angular momentum will
be modified even if the star has {{{}}}$J= J_{\rm LSO}$. As a result
it is very likely that the star will be scattered into the loss-cone
(or away from it, to an orbit where energy dissipation is very
weak). The approximate condition $t_0(a, J\!\to\! J_{\rm
LSO})\!<\!t_{lc}(a)$ translates into a maximal semi-major axis $a_{\rm
GW}$ a CR must have in order to spiral in without plunging into the
MBH, and become a \LISA source. For MBHs, $a_{\rm GW}={\rm
few}\times0.01\pc$
\citep{Hop05, Hop06,Hop06b}.

In order to test these ideas further, Hopman \& Alexander
\citep{Hop05} performed Monte Carlo (MC) simulations. In these
simulations, stars started at some given energy and angular momentum,
and at every time-step $\delta t$, the angular momentum changed
randomly by an amount $\delta J=(\delta t/t_r)^{1/2}J_c$, and the
energy was increased deterministically due to GWs by an amount $\delta
E=\dot{E}_{\rm GW}\delta t$. For each energy, the experiment was
performed many times with different random numbers, and for each star
it was denoted whether it entered the \LISA band, or whether it was
swallowed prematurely by the MBH because the angular momentum became
smaller than that of the last stable orbit. The orbital quantities in
these simulations were treated in a pseudo general relativistic
form. If $N_p(a)$ is the number of stars with some given initial $a$
that are promptly swallowed, and $N_i(a)$ the number of stars which
spiral in slowly and become sources that can be observed with \LISA,
then the probability for inspiral is given by

\begin{equation}\label{e:Sa}
S(a)\equiv {N_i(a)\over N_p(a) + N_i(a)}.
\end{equation}

\begin{figure}
\includegraphics[angle=0,scale=.37]{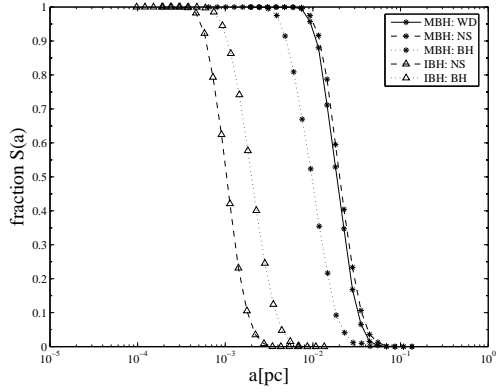} 
\caption{Dependence of the normalized
fraction of inspiral events on initial semi-major axis, for the case
of a MBH of $\Mbh\!=\!3\times10^{6}\,\Mo$ (asterisks) and an IMBH of
$\Mbh\!=\!1\times10^{3}\,\Mo$ (triangles). The solid line is for WDs,
the dashed lines for NSs, and the dotted lines for stellar
BHs. Reprinted with permission from the Astrophysical
Journal.\label{f:Sa}}
\end{figure}

In figure (\ref{f:Sa}) the resulting function $S(a)$ from the
simulations performed by \citep{Hop05} is shown. The left two curves
are the results for IMBHs and the three right curves for a MBH of mass
$\Mbh=3\times10^6\Mo$. The function $S(a)$ is equal to 1 for small
$a<a_{\rm GW}$, and drops very quickly to $S(a)=0$ for $a>a_{\rm GW}$,
in accordance with the qualitative picture given above. To account for
the fact that not all stars that are captured become eventually
observable by \LISA, the inspiral rate is obtained by integrating
equation (\ref{e:Flow}) with $S(a)$,

\begin{equation}\label{e:Gam}
\Gamma =
\int_{0}^{\infty}daS(a)\frac{N(a)}{\ln(J_{c}/J_{lc})t_{r}(a)}\,.
\end{equation}

The shape of $S(a)$ was verified by $N$-body simulations
\citep{Bau05b} for tidal inspiral of MS stars near an IMBH, a process
that is dynamically similar to GW dissipation
\citep{Ale03a, Hop04, Hop05b}.

\begin{figure}
\begin{minipage}[b]{0.5\linewidth} 
\centering \includegraphics[width=6cm]{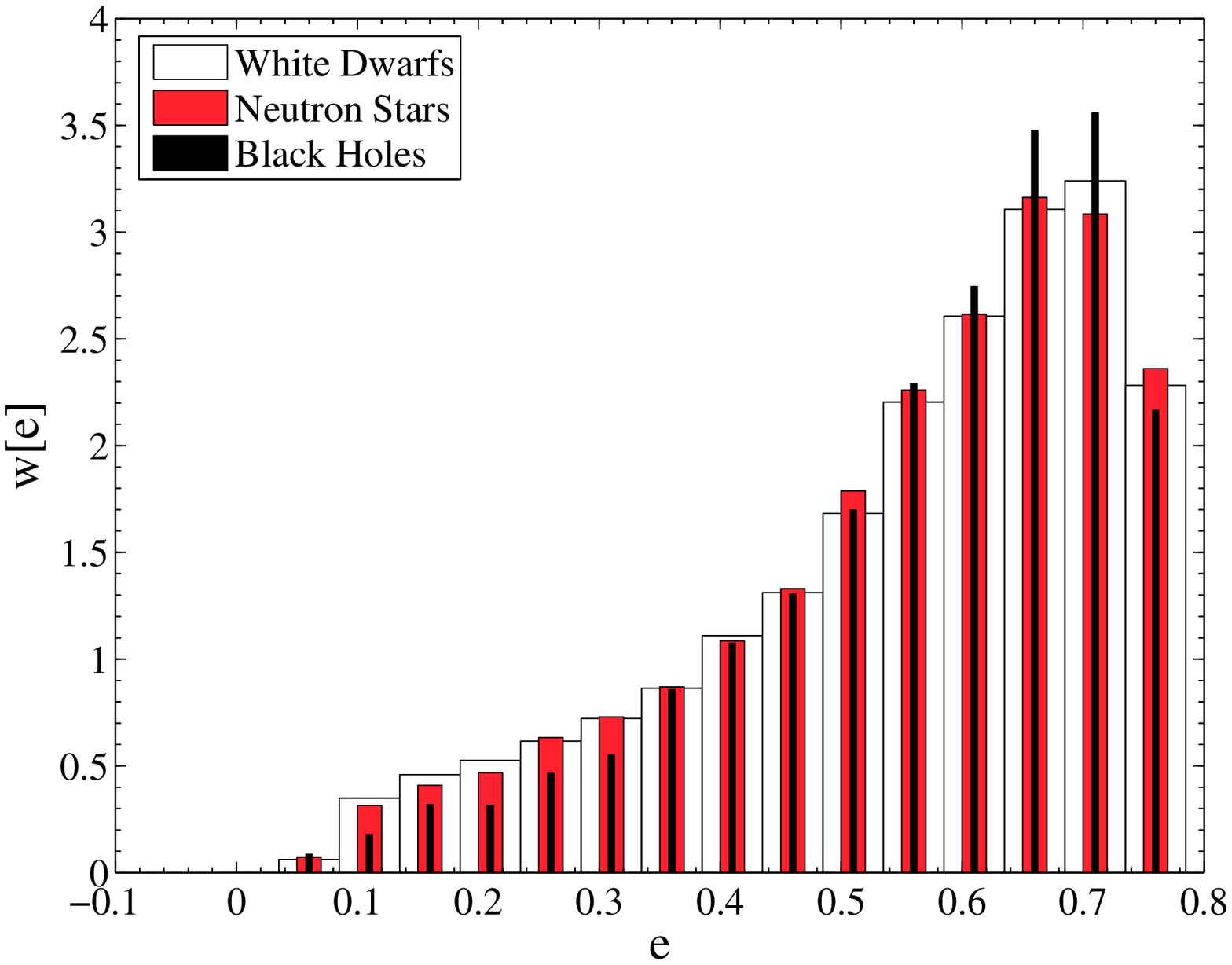}
\caption{Eccentricity DF for orbits of $P=10^4\, {\rm s}$, from the MC simulations by \citep{Hop05}. The left figure is for MBHs with $\Mbh=3\times10^6\Mo$, and the right figure for IMBHs with mass $\Mbh=1\times10^3\Mo$. The eccentricities for IMBHs are so high, that the frequency may be too high to be observed by \LISA. Reprinted with permission from the Astrophysical Journal.}
\end{minipage}
\hspace{0.5cm} 
\begin{minipage}[b]{0.5\linewidth}
\centering
\includegraphics[width=6cm]{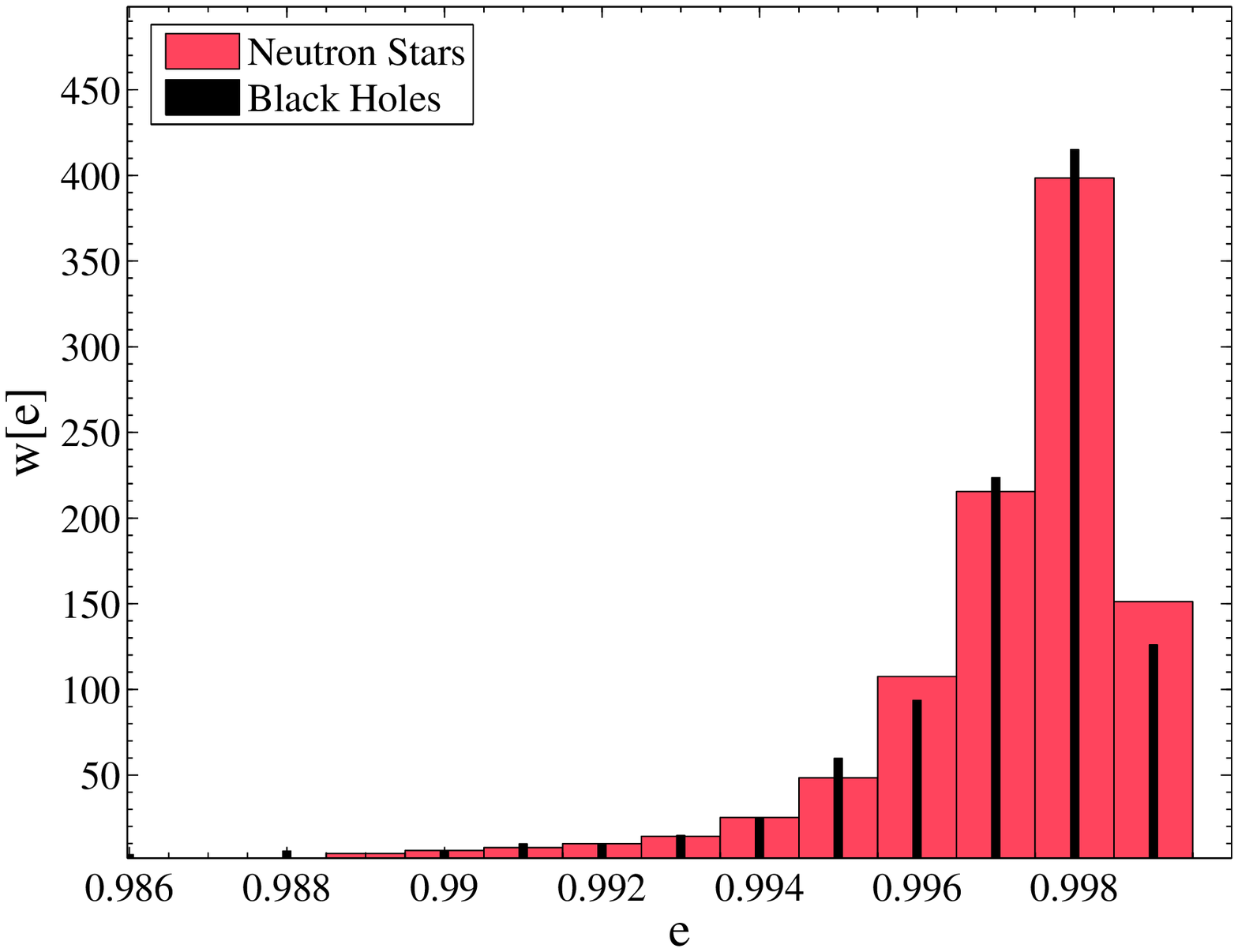}
\caption{Eccentricity DF for orbits of $P=10^4\, {\rm s}$, from the MC simulations by \citep{Hop05}. The left figure is for MBHs with $\Mbh=3\times10^6\Mo$, and the right figure for IMBHs with mass $\Mbh=1\times10^3\Mo$. The eccentricities for IMBHs are so high, that the frequency may be too high to be observed by \LISA. Reprinted with permission from the Astrophysical Journal.\label{f:ecc}}
\end{minipage}
\end{figure}

For successful inspirals in the MC simulations by \citep{Hop05}, the
eccentricities were recorded at the point where the period was
$P=10^4\, {\rm s}$. The DF of eccentricities was seen to be skewed
towards high eccentricities (Fig. \ref{f:ecc}). Interestingly, this is
in contrast with what is predicted for EMRIs that result from {\it
indirect} capture of stars (see below).

%
%
%
%

\section{Dynamics within the inner few $0.01\pc$}\label{s:0.01}

The EMRI event rate is determined by the dynamics and stellar content
within $a_{\rm GW}={\rm few}\times0.01\pc$, which determine $N(a)$ and
$t_r(a)$ in Eq. (\ref{e:Gam}). In this section, two important
dynamical phenomena are discussed.

\subsection{Mass segregation}\label{s:massseg}

When two stars of different masses interact gravitationally, there
will be a tendency to divide their kinetic energy equally. While
equipartition is never reached for a star cluster, this effect causes
the massive stars to sink more effectively to the center than the
lighter stars, and is known as mass segregation. Mass segregation near
a MBH was first studied by \citep{Bah77} in the context of an old
stellar population near an IMBH. They found that different species are
distributed according to approximate power-laws with different slopes
(see Eq. \ref{e:cusp}), and that the most massive particles have the
steepest slopes and are thus more centrally concentrated. This effect
was confirmed by $N$-body simulations by \citep{Bau04b, Fre06}, whose
results were generally in good agreement with the \citep{Bah77}
predictions. Simulations of mass segregation near MBHs were recently
performed by \citep{Fre06} and \citep{Hop06b}. Freitag et
al. \citep{Fre06} performed extensive MC simulations \citep{Fre02}
with different mass functions that included much larger mass ratios
for the different stellar populations than were considered by
\cite{Bah77}. Their simulations confirm the results by \cite{Bah77}
that the massive species are steeper distributed than the lighter
stars, although they find that the cusp of the lightest stars is even
shallower ($\alpha=1.3$), and the more massive stars is steeper
($\alpha\sim2$) than those found by
\citep{Bah77}.

\begin{figure}
\includegraphics[angle=0,scale=.37]{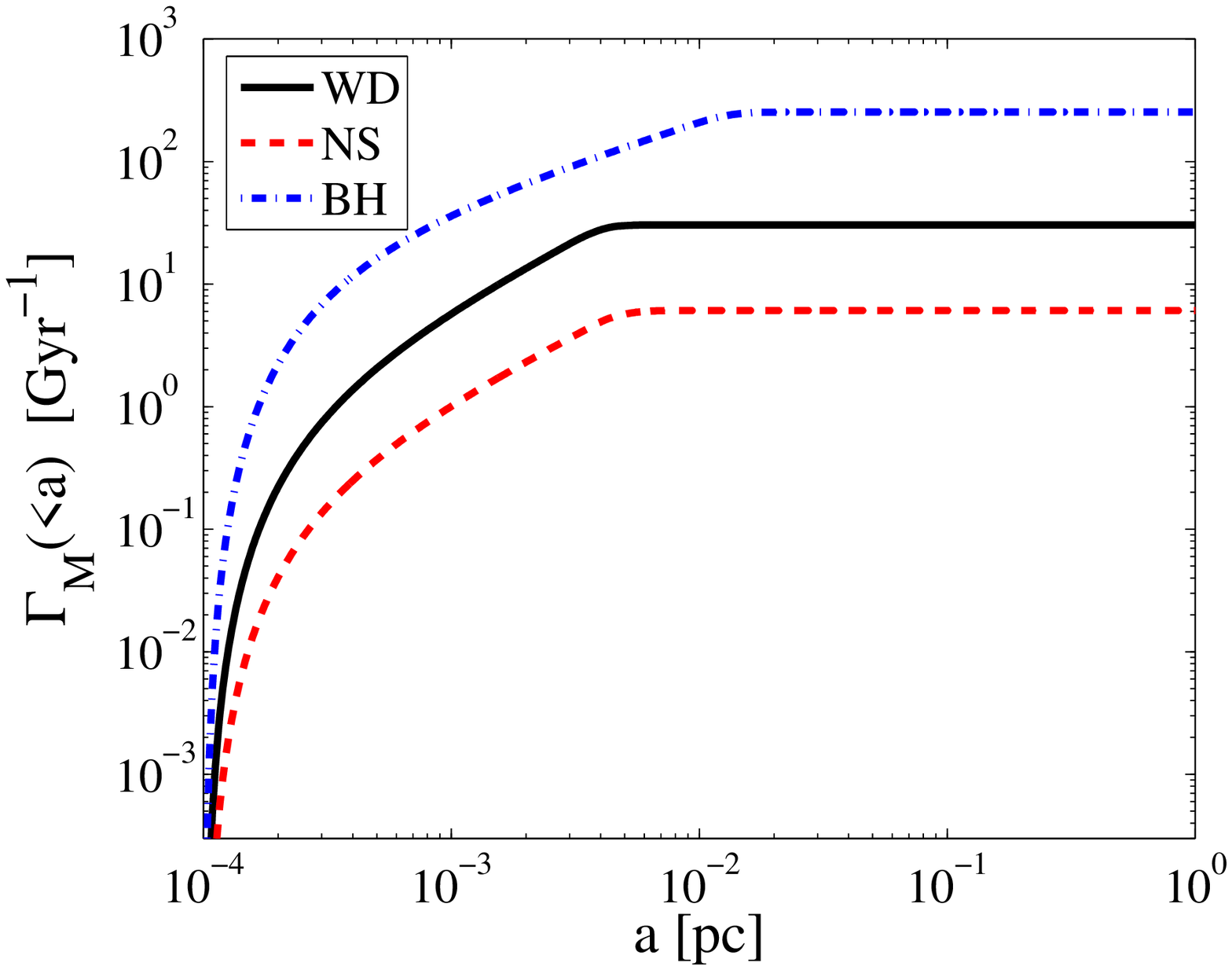}
\caption{Cumulative rate of EMRI formation for stellar BHs (blue dashed-dotted line), WDs (black solid line), and NSs (red dashed line) as a function of the distance from the MBH. In spite of the fact that far away from the MBH WDs were assumed to be more abundant than stellar BHs by two orders of magnitude, stellar BHs dominate the even rate. Reprinted with permission from the Astrophysical
Journal.\label{f:cumcap}}
\end{figure}

Hopman \& Alexander \citep{Hop06b} directly solved the multi-mass
Fokker-Planck equations given by \cite{Bah77} for four species: MSs,
WDs, NSs and stellar BHs. The equations are one dimensional and
consider only the dynamics in energy space, but they added a sink term
to calculate the rate at which stars reach the loss-cone in
$J$-space. Like
\citep{Fre06}, they find that the steepest DF is that of the stellar
BHs, with a slope of {{{}}}$\alpha=2$. Within a distance of
{{{}}}$<0.1\pc$ of the MBH stellar BHs become the dominant CR in terms
of numbers. Since the stellar BHs are more massive than the other
stars, this also implies that the relaxation time decreases when
compared to a single mass population;
\cite{Hop06b} find that close to the MBH it becomes as short as $\sim 10^8\yr$. 
Using the relaxation time resulting from the steady state DF of the
stars, they determined the probability function for successful
inspiral, Eq. (\ref{e:Sa}), with the MC method described in
\cite{Hop05}. Finally, now that both the radial DF of
stars and the inspiral probability function are known, the cumulative
event rate can be estimated for the different species with
Eq. (\ref{e:Gam}); the result for WDs, NSs and stellar BHs are given
in Fig. (\ref{f:cumcap}). The total rate is clearly dominated by
stellar BHs, and estimated to be of order $\Gamma={\rm
few}\times10^2\perGyr$ per MBH system. Since stellar BHs also give a
stronger signal due to their higher mass, it appears a robust result
that the EMRI detection rate by \LISA will be dominated by stellar
BHs.

The DF of stars within the radius of influence depends not only on
stellar dynamics, but also on poorly constraint quantities such as the
star formation history and initial mass function outside the radius of
influence $r_h$
\citep{deF06}. Even in our own Galactic center, there is no
knowledge of the stellar content within $a_{\rm GW}$ \citep{Mou05}.
EMRI observations by \LISA may therefore provide a unique observation
to the stellar populations very close to MBHs.

\subsection{Resonant relaxation}\label{s:rr}

Near MBHs orbits of stars hardly precess if the enclosed mass in stars
is small, and precession due to general relativistic effects can be
neglected. It was shown \citep{Rau96, Rau98} that the resulting
torques for such orbits lead to a relaxation mechanism that is much
more efficient than two-boy relaxation. This mechanism, known as {\it
resonant relaxation} (RR), affects only the angular momenta of the
stars, and not their energies.

If general relativistic effects are neglected, the RR time $T_{\rm
RR}$ is given by
\begin{equation}
T_{\rm RR}=A_{\rm RR}{\Mbh\over\Ms}P(a);
\end{equation}
here $A_{\rm RR}$ is a numerical factor which was estimated by
\citep{Rau96} to equal $A_{\rm RR}=3.56$. Since $T_{\rm RR}\propto P(a)$, the RR time decreases rapidly towards the MBH. 

RR leads to a small enhancement of the tidal disruption rate, which is
dominated by effects close to the radius of influence ($r_h\sim1\pc$),
where $T_{\rm RR}$ approximately equals the non-resonant relaxation
time. However, at the critical distance $a_{\rm GW}\sim0.01\pc$, RR is
very efficient.  As a result, the flow of stars towards the loss-cone
is much more rapid than would be the case if relaxation were be
non-resonant, as is usually assumed. Hopman \& Alexander \citep{Hop06}
showed that the EMRI event rate is in fact dominated by RR, which
increases the rate by nearly an order of magnitude.

The factor $A_{\rm RR}$ is not very well determined. If it is slightly
larger than estimated by \citep{Rau96}, the GW event rate will
increase further. However, if $A_{\rm RR}$ is larger by an order of
magnitude than estimated by \citep{Rau96}, the loss-rate can be so
large, that energy relaxation cannot replace the stars, which can in
principle lead to a strong depression of the event rate
\citep{Hop06}. It is therefore of interest to obtain further
understanding of the details of RR.

\section{Indirect capture of EMRIs}\label{s:indirect}

In the scenarios described above, a CR on a loosely bound orbit
becomes tighter bound because it loses energy to the GWs
themselves. This happens on extremely eccentric orbits
($1-e\sim10^{-4}$), and even after the orbital frequency exceeds
$10^4{\,\rm Hz}$, the orbit is still eccentric
(Fig. \ref{f:ecc}). There are several alternatives in which the CR
does not lose its energy to GWs, but to other channels. Here three
such mechanisms are discussed. Interestingly, these all have in common
that the resulting eccentricities become very low, $1-e\sim1$.

\subsection{Stellar formation in accretion disks}\label{s:acc}

When dense gas is present near a MBH in the form of an accretion disk,
as is the case for active galactic nuclei, it may become
self-gravitating, and stars can form in the disk. Our own Galactic
center provides evidence that this has happened there recently
($<{\rm few}\,\Myr$), because there are disks of young OB stars present
\citep{Lev03, Gen03a}, which are best explained by stellar formation in
an accretion disk \citep{Lev03, Lev06, Nay05c}.  Levin
\citep{Lev03a} argued that massive stars formed in accretion disks may
have enough time to evolve and become stellar BHs, and spiral in due
to interaction with the gas in the disk within the life time of the
accretion disk. The interaction with the gas flow is likely to keep
the orbit close to circular, so this would lead to zero-eccentricity
\LISA EMRIs. The event rate depends clearly on a large number of
uncertain factors, such as the efficiency of star formation in disks,
the accretion rate of the stars in the disks, and the inspiral time
within the disk. However, \citep{Lev03a, Lev06} found that it is not
implausible that the contribution to the total
\LISA event rate may be considerable.

\subsection{Tidal binary disruption}

The majority of stars in galaxies reside in binaries. Near MBHs the
binary fraction may be considerably lower than average, because of the
high velocity dispersion: this implies that typically, when a single
star encounters a binary, the binary star becomes less tight
\citep{Heg75}. Nevertheless, it is plausible that near the radius of
influence of a MBH there is a binary fraction of at least a few
percent \citep{Per06}.

When a binary is scattered to a very eccentric orbit, it may be
tidally disrupted by the MBH. In that case, one star is ejected at
very high velocity, and the other star becomes tightly bound to the
MBH
\citep{Hil88, YuQ03, Gua05, Per06}. In this case it is the ejected star that
'absorbs' the orbital energy; such hypervelocity stars have now been
observed in our Galaxy \citep{Bro05}.

If the star that remains bound to the MBH is a CR, it can spiral in
due to GW emission. Miller et al. \cite{Mil05} found that this results
in zero-eccentricity \LISA orbits, and that the event rate may well
exceed the event rate of direct inspirals, although uncertainties are
considerable, in particular due to the unknown fraction of binaries
near MBHs, and the frequencies in which CRs are present in such
binaries.

\subsection{Remnants of ultraluminous X-ray sources}

It was argued by \citep{Hop05} that for direct capture of a CR by an
IMBH, the resulting orbits will be so eccentric that even if a star is
captured, it will not be 'visible' by \LISA because the frequency will
be too high. However, this does not necessarily imply that IMBHs will
not be observable at all by \LISA. A possibility that was considered
by \cite{Hop05b} is that a main sequence star that is {\it tidally}
captured by an IMBH will evolve later to a stellar BH and spiral in
due to GW emission. Since the tidal forces have circularized the
orbit, this will lead to a low eccentricity and low frequency orbit
that can be observed by \LISA. Before the super nova explosion, the
star may fill its Roche lobe and be observable as an ultraluminous
X-ray source \cite{Hop04}. Observational support for this scenario
comes from the ultraluminous X-ray source in M82, which is perhaps the
best IMBH candidate. This source has a 62 day period in its X-ray
luminosity, which has been interpreted as an orbital period, and is
best explained if the accretor is an IMBH \citep{Kaa06, Pat06}.

As for the previous scenarios, it is highly uncertain how many of
these sources will be observable by \LISA (if any). \citep{Hop05b}
estimated from the observed ultraluminous X-ray populations that there
is a good probability to observe several IMBH-CR binaries with \LISA.

\section{Conclusions}
I have discussed the formation of EMRI sources that will be observable
by \LISA. The event rate for directly captured EMRIs is of the order
of ${\rm few}\times10^{2}\perGyr$ per MBH, with stellar BHs the dominant
species. This event rate will lead to many observable sources for
\LISA \citep{deF06}. EMRIs formed by direct capture are typically highly
eccentric. Alternative mechanisms for EMRI formation lead to very low
eccentricities, so that the resulting eccentricity DF will probably be
bimodal.

\begin{theacknowledgments}
I thank the organizers of the LISA6 conference for a very stimulating meeting.
\end{theacknowledgments}


\end{document}